\documentclass[epj, onecolumn]{webofc}
\usepackage[varg]{txfonts}   
\woctitle{ARENA}
\begin{document}
\title{In-situ absolute calibration of electric-field amplitude measurements with the LPDA radio detector stations of the Pierre Auger Observatory}
%
%

\author{\firstname{Florian} \lastname{Briechle}\inst{1}\fnsep\thanks{\email{briechle@physik.rwth-aachen.de}} for the Pierre Auger Collaboration\inst{2}}

\institute{RWTH Aachen University, III. Physikalisches Insitut A
\and
Full author list: \url{http://www.auger.org/archive/authors_2016_06.html}}

\abstract{%
 With the Auger Engineering Radio Array (AERA) located at the Pierre
Auger Observatory, radio emission of extensive air showers is observed.
To exploit the physics potential of AERA, electric-field amplitude measurements with the radio detector stations need to be well-calibrated on an absolute level. A convenient tool for far-field calibration campaigns is a flying drone. Here we make use of an octocopter to place a calibrated
source at freely chosen positions above the radio detector array. Special emphasis is put on the
reconstruction of the octocopter position and its accuracy during the
flights.

The antenna response pattern of the radio detector stations
was measured in a recent calibration campaign. Results of these measurements are presented and compared to simulations. It is found that measurements and simulations are in good agreement.

}
\maketitle
\section{Introduction}
\label{intro}


To do high precision measurements of ultra-high energy cosmic rays with the Auger Engineering Radio Array (AERA) \cite{SchulzProcICRC2015, GlaserArena2016} at the Pierre Auger Observatory \cite{Aab_OverviewAuger15}, the radio detector needs to be well-calibrated. The largest challenge in performing such a calibration is to place a signal source at every direction relative to the detector station and  at least 20\,m away to guarantee far-field conditions. Different calibration methods have been developed and already performed at AERA \cite{Abreu_aeracalib2012} and other experiments \cite{2015ApelLOPES_improvedCalibration, TunkaRex_NIM_2015, NellesLOFAR_calibration2015}. In this contribution, we present a novel and very flexible method using a flying drone to carry the signal source. A drone is a very useful tool, since it can be placed at any point above the array and can carry any piece of equipment as long as it is not too heavy. With the presented method, a higher accuracy can be achieved than in the previous calibrations mostly due to a larger set of measurements and a better handling of environmental conditions.


%
At AERA, two different antenna types are used. The core consists of a dense array of 24 stations equipped with log-periodic dipole antennas (LPDA). They consist of nine separate dipoles optimized for a frequency range of 30 to 80 MHz. The other stations are equipped with butterfly antennas. However, in this proceeding, only LPDA stations are considered.

Extensive air showers produce short radio pulses. These pulses then induce a signal measurable as voltages at the antenna output. The value which relates the Fourier transformed electric field $\mathcal{\vec{E}}$ to the measured Fourier transformed voltage $\mathcal{U}$ is called the vector effective length (VEL) $\vec{H}$:

\begin{equation}
 \mathcal{U}(f, \theta, \phi) =\vec{H}(f, \theta, \phi) \cdot \mathcal{\vec{E}}(f).
\end{equation}

$\vec{H}$ depends on the frequency of the signal $f$ and the incoming direction, described by the zenith angle $\theta$ and the azimuth angle $\phi$. In spherical coordinates, $\vec{H}$ is a superposition of a horizontal $H_\phi$ and a perpendicular meridional $H_\theta$ component. The absolute value of these components is determined in a transmission measurement using a calibrated transmitting antenna. For the absolute value of a VEL component at a specific frequency and direction the following holds

\begin{equation}
 |H_{i}| = \sqrt{\frac{4 \cdot \pi \cdot Z_A}{Z_0}} \cdot R \cdot \sqrt{\frac{P_{r, i}}{G_t \cdot P_t}},
 \label{eq:absH}
\end{equation}

where $i = \theta, \phi$ denotes the VEL component, $Z_A$ denotes the impedance of the antenna, \hbox{$Z_0 = 120 \,\pi\,\Omega$} is the vacuum impedance, $R$ describes the distance between the antenna under test (AUT) and the transmitting antenna, $P_r$ denotes the power received at the AUT, and $G_t$ and $P_t$ are the gain and power of the transmitting antenna \cite{WeidenhauptPHD2015}.

\section{Calibration Setup}

\begin{figure}
\centering
\begin{minipage}{.5\textwidth}
  \centering
  \includegraphics[width=0.9\linewidth]{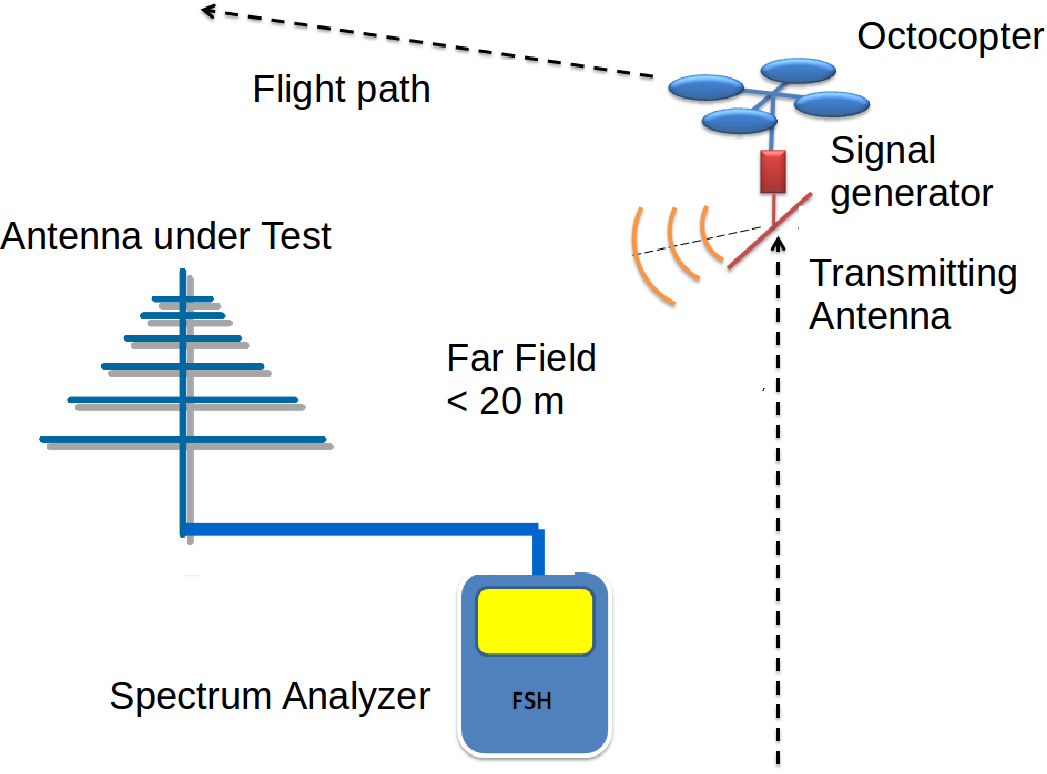}  
\end{minipage}%
\begin{minipage}{.5\textwidth}
  \centering
  \includegraphics[width=.9\linewidth]{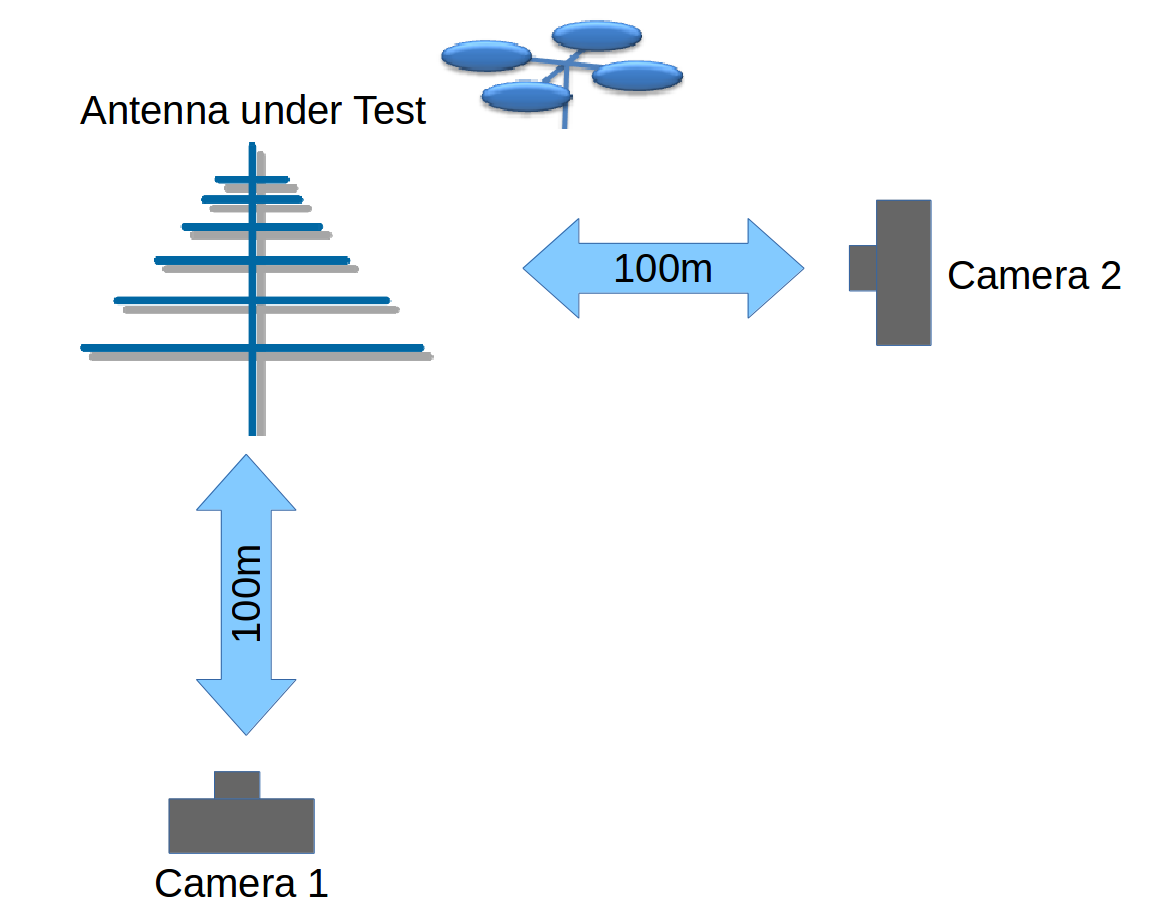}  
\end{minipage}
  \begin{minipage}[t]{.4\linewidth}
    \caption{Experimental setup for the calibration  measurement.}
    \label{pic:setup}
  \end{minipage}%
  \hspace{1cm}
  \begin{minipage}[t]{.4\linewidth}
    \caption{Experimental setup for the optical position determination.}
    \label{pic:setupcam}
  \end{minipage}
\end{figure}

A sketch of the setup for the calibration measurements is shown in Fig~\ref{pic:setup}. The AUT is connected to a spectrum analyzer model Rohde \& Schwarz FSH 4. This device measures the received power at each frequency. A signal generator and a calibrated transmitting antenna are attached to an octocopter and placed at a distance of at least 20\,m to make sure far-field conditions apply. The signal generator RSG-1000 produces continuous signals at every 5 MHz. During the measurement the octocopter then flies straight up to a height of approximately 30\,m, flies towards the antenna until it is directly above it. Afterwards, it flies back and finally lands at the starting position. In this way, the complete zenith angle range is covered.

\section{Reconstruction of the Octocopter Position}

As seen in Eq.~\eqref{eq:absH}, the distance $R$ between the transmitting antenna used for the calibration and the AUT is of crucial importance. The positions of the antennas at AERA have been previously measured with a differential GPS system and are known to within a few cm. For the position of the octocopter, an internal GPS receiver for the position in the x-y plane and a barometer for the height measurement are built-in. The statistical uncertainty of the position measurement has been found to be better \hbox{than 1\,m}. However, the systematical uncertainty is on the order of a few meters. To make a high precision measurement with small systematical uncertainties, another position determination method has been developed. It makes use of two standard digital cameras. The setup is shown in Fig~\ref{pic:setupcam}. Two cameras are placed on orthogonal axes with a distance of $\sim$100m to the AUT. Canon Ixus 132 cameras with a resolution of 16 Megapixel are utilized. They are set to an autonomous mode where they take pictures every three seconds. 

From these pictures the position of the octocopter during the full flight is reconstructed. In the first step, the pixel tuple of the octocopter in each picture is determined by a ``template matching'' algorithm. Using this pixel tuple, a vector pointing from each camera to the octocopter is constructed. The point of closest approach of these vectors is calculated and used as the reconstructed octocopter position. For more details on the whole reconstruction process, see \cite{BriechleMa2015}.


From comparison with a differential GPS and the built-in position sensors, a statistical uncertainty of 1.5\,m and a systematical uncertainty of 1\,m for this optical method has been found. So, the optical method compared to the built-in sensors has a smaller systematical uncertainty, albeit with a larger statistical uncertainty.  Therefore, in the calibration flights the built-in sensors are used, corrected by the optical method. In the x-y plane, the  absolute offset between the built-in sensors and the optical method, measured in each separate flight, is added, while in the height, the height measured with the built-in sensors is multiplied by the ratio of the measured height with the built-in sensors and the height measured with the optical method.

\section{Results of the Calibration Measurements}

\begin{figure}
\centering
\begin{minipage}{.5\textwidth}
  \centering
  \includegraphics[width=0.9\linewidth]{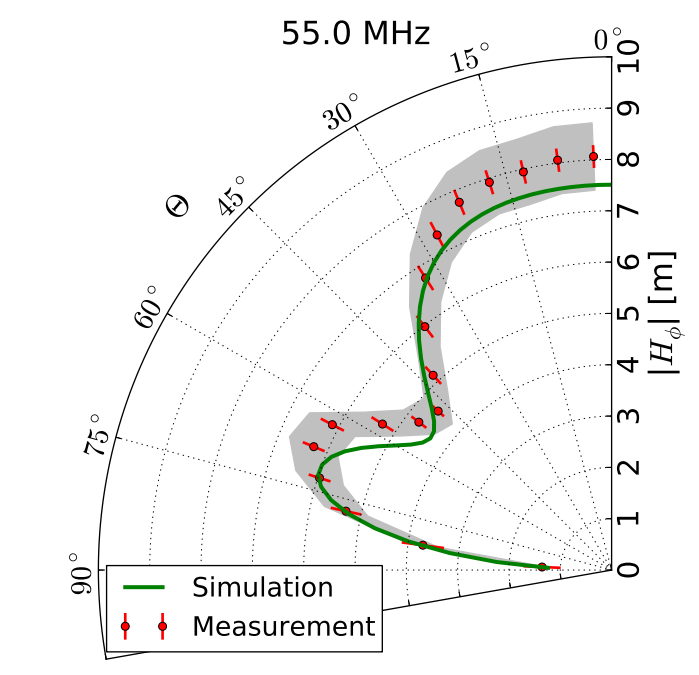}  
\end{minipage}%
\begin{minipage}{.5\textwidth}
  \centering
  \includegraphics[width=.9\linewidth]{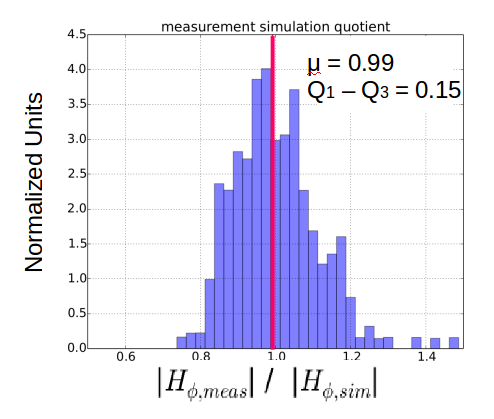}  
  
\end{minipage}
  \begin{minipage}[t]{.41\linewidth}
    \caption{$|H_{\phi}$| at 55 MHz as function of $\theta$. The red points denote the measured values, the red error bars the statistical uncertainty and the grey color band the systematical uncertainty. The green line denotes simulated values.}
    \label{fig:ex} 
  \end{minipage}%
  \hspace{2cm}
  \begin{minipage}[t]{.41\linewidth}
    \caption{Ratio of measured and simulated $|H_{\phi}$|. Each entry represents one specific frequency and zenith angle. Q1 and Q3 denote the 25\% and 75\% quartiles, respectively.}
    \label{fig:simquothist} 
  \end{minipage}
\end{figure}

Several flights for measuring both $|H_{\phi}$| and $|H_{\theta}$| over a time of a few weeks have been performed. In Fig.~\ref{fig:ex}, the results of the calibration for  $|H_{\phi}$| as a function of $\theta$ at a frequency of 55\,MHz for a fixed azimuth angle $\phi = 0^\circ$ are shown. Each value represents the mean over a zenith angle \hbox{range of 5$^\circ$.} The error bars denote the statistical spread of the measurement while the error bands denote the systematical uncertainties. These uncertainties are dominated by the spectrum analyzer, which has an uncertainty of 0.5\,dB given by the manufacturer. This uncertainty has to be applied to the injected and the received power. For a frequency of 55\,MHz, which is the middle of the frequency spectrum of the antennas, and a zenith angle of $45^\circ$, from which most of the air shower signals originate, a statistical uncertainty of 4.4\% is found. The systematical uncertainty is found to be 8.1\%. These values do not vary much over the whole frequency and directional range. Combining these uncertainties gives an overall uncertainty of 9.3\%. This is a significant improvement on the previous calibration, which had an uncertainty of 12.5\% \cite{Abreu_aeracalib2012}. 

The green line in Fig.~\ref{fig:ex} shows a simulation of the antenna pattern done with the 4NEC2 simulation code. In Fig.~\ref{fig:simquothist}, the ratios of the simulated and measured VEl for each 5$^\circ$ zenith angle bin and each measured frequency are histogrammed. The mean of the distribution is 0.99, which shows a good agreement between simulation and measurement.

The same analysis has also been done for the meridional component $|H_{\theta}|$. It is found that the sensitivity is much lower than in the horizontal component.

\section{Summary and Outlook}

A calibration campaign of the LPDA antennas at AERA using an octocopter and a calibrated reference source has been performed. To determine the position of the reference source during the flight, a new optical reconstruction method has been developed and evaluated. The absolute vector effective length, which describes the antenna response pattern to an incoming electric field, has been measured with an uncertainty of 9.3\%. A good agreement between simulation and measurement was found. A journal article about the whole calibration process is currently in the works.

This calibration is a crucial component in the determination of an absolute energy scale of cosmic rays from first principles. Using the new calibration presented here, the uncertainty of the cosmic-ray energy measurements can be lowered with regard to current values \cite{Aab_energyaera16, Aab_energyaera_prd2}.  In the future, these results from AERA will be used to improve the measurements of the whole Pierre Auger Observatory \cite{GlaserArena2016}.

\bibliography{generalbib.bib}

\end{document}